\documentclass[preprint,12pt]{elsarticle}

\usepackage{amssymb}

\usepackage[version=3]{mhchem}
\usepackage[]{natbib}
\usepackage{graphicx}
\usepackage{hyperref}

\journal{Solid State Communications}

\begin{document}

\begin{frontmatter}
\biboptions{sort&compress}



\title{Excitons in \ce{Mg(OH)2} and \ce{Ca(OH)2} from \textit{ab initio} calculations}


\author{A. Pishtshev}

\address{Institute of Physics, University of Tartu, 51014 Tartu, Estonia}

\author{S. Zh. Karazhanov}
\address{Department for Solar Energy, Institute for Energy Technology, NO-2027 Kjeller, Norway}
\ead{smagulk@ife.no}
\author{M. Klopov}

\address{Department of Physics, Tallinn University of Technology, 19086 Tallinn, Estonia}

\date{\today}

\begin{abstract}
By using \textit{ab initio} calculations with the HSE06 hybrid functional and GW approximation combined
with numerical solution of the Bethe Salpeter equation (GW-BSE) we predict the existence of diverse number
of excitonic states in multifunctional hydroxides \ce{\textit{X}(OH)2} (\textit{X}= Mg and Ca) that were
not previously reported experimentally or theoretically. Imaginary part of the dielectric function and
reflectivity spectra show very strong peaks corresponding to the electron-hole pair states of large
binding energy. The origin of the excitons is attributed to strong localization of the hole and electron
associated to oxygen $2p_x, 2p_y$ occupied states as well as to oxygen and earth metal $s$ empty states,
respectively. The results have important implications for different applications of the materials in
optoelectronic devices.
\end{abstract}


\end{frontmatter}



\section{Introduction}
\label{Intro}

Electronic many-body effects play a key role in the electrical and optical properties of solids
\cite{Louie1998,Louie2000}. In particular, the proper accounting for two-particles excitation energies
allows one to explain specific features of the optical absorption spectrum of semiconductors and
insulators \cite{Menke2014,Koch2006,Scholes2006}. Due to specific features of dielectric screening, e.g.,
low dielectric constants, the interaction of the excited electrons and holes differs significantly in
wide-gap insulators than that in semiconductors. Exciton binding energy resulting from the strong $e-h$
interaction is often of large magnitude \cite{Louie2000}. In this context, the main focus of the present
work is the principal case study demonstrating that the strength of the $e-h$ interaction can be modulated
by materials engineering methods. For example, the experimentally established exciton binding energy is
86~meV for \ce{MgO} and 95~meV for \ce{CaO} with band gap of 7.775~eV and 7.034~eV \cite{Whited1969},
respectively: both materials possessing predominantly ionic type of chemical bonding. By means of
hydration of these materials the existing ionic bonds are reorganized so that a certain portion of
covalency is supplied by the water molecules \ce{H2O}. The resulting crystalline alkali-earth hydroxides
\ce{\textit{X}(OH)2} (\textit{X}=Mg, Ca) acquire a system integrity that is governed by the negative
oxygen ions via a bridging combination of the strong covalent bonding in the \ce{OH^{-}} hydroxyl anions
and \ce{\textit{X}-O} strong ionic connections \cite{ASM2014_1}. This in turn determines a
multifunctionality property of \ce{\textit{X}(OH)2} compounds, which is generally based on the equal
utilization of electronic characteristics such as large band gap and low refractive index (relevant to
purely ionic systems) and covalent contributions from the oxygen within a common crystal-chemical
environment. Our preliminary considerations applied to \ce{\textit{X}(OH)2}  have shown that this feature
contributes significantly to the nature of optoelectronic properties, so that one can expect in these
materials strong Coulomb attraction between an excited electron and a hole and, consequently, generation
of an exciton with a large binding energy.
Excitonic contributions to the optical absorption process can be described by numerical solution of the
Bethe-Salpeter equation (BSE) with high accuracy
\cite{Louie1998,Bohn_excitons_1998,Albrecht_excitons_1998,Louie_MoS2_2013,Su-Huai_exciton_2013}. In this
work we will study the electronic structure and optical properties of \ce{\textit{X}(OH)2} and report
about band gap excitons with large binding energy of 0.46~eV for \ce{Mg(OH)2} and 0.85~eV \ce{Ca(OH)2}. We
identify and analyze the character of exciton peaks in the optical spectrum and relate them to strong
localization of the hole and electron to oxygen $2p_x, 2p_y$ occupied states as well as to oxygen and
metal $s$ empty states, respectively. The fundamental aspect of the article is that we have found that in
crystalline hydroxides \ce{Mg(OH)2} and \ce{Ca(OH)2} the many-body effects in the optical absorption
spectra play a crucial role. From a practical point of view, materials under consideration may serve as
interesting hosts, so that our results definitely have important implication for different applications of
the materials in optoelectronic devices \cite{Miyazaki2006,Yum2006,Huang2011}.

\section{Structural model and computational details}
\label{methods}

The multifunctional hydroxides \ce{\textit{X}(OH)2} crystallize into the \ce{CdI2}-type structure of space
group of $D^3_{3d}$(Refs.~\cite{Desgranges1996,Busing1957,Desgran1993}) with the unit cell containing two
metal cations and two \ce{OH-} anionic groups. We have employed Vienna \textit{ab initio} simulation
package (VASP) \cite{VASP1996} together with the potential projector augmented-wave (PAW) method
\cite{Bloechl1994,Joubert1999,Hafner1993}. A ${\Gamma}$-centered optimized $8{\times}8{\times}8$ ${\bf
k}$-point mesh was selected for all DFT and GW calculations. To ensure data accuracy and clarity, the
calculations were performed within the large plane-wave basis set with $800$~eV cutoff, and with the
application of GW-versions of PAW Perdew-Burke-Ernzerhof (PBE) pseudopotentials \cite{Ernzefhof1996},
which represent $2p^6 3s^2$, $3s^2 3p^6 4s^2 3d^0$, $2s^2 2p^4$, and $1s^1$ valence electron
configurations for \ce{Mg}, \ce{Ca}, \ce{O} and \ce{H} atoms, respectively. In our computations spin-orbit
coupling was not included into consideration. Studies of ground state properties and electronic structure
have been performed within the Heyd-Scuseria-Ernzerhof (HSE06) hybrid functional
\cite{Scuseria2003,Scuseria2006,Scuseria2010}. Optical properties have been studied by using the many-body
Hedin's GW approximation \cite{Hedin1965} combined with numerical solution of BSE according to the
strategy developed in Refs.~\cite{Kresse2007GW_2,Marom2012,Friedrich2012}, which utilizes the wave
functions derived from hybrid functional calculations as a starting point for subsequent GW numerical
procedures. For description of exchange effects in a periodic ion-covalent insulating system we employ the
relation $a{\approx}{\epsilon^{-1}_{\infty}}$
(Ref.~\cite{Alkauskas2008,Botti2010,Alkauskas2010,Botti2011,Franchini2012}) that approximates the mixing
coefficient $a$ (fraction of the Fock exchange) in the xc potential in terms of the effective screening of
the bare nonlocal exchange part of the electron-electron interaction \cite{Botti2010,Botti2013}.

\section{Results}
\label{results}

 In our study, we start with preliminary series of calculations in order to
ascertain the relevant value of $a$ via ${\epsilon^{-1}_{\infty}}$ that will consistently be utilized
further as a material-specific parameter related to the short-range Fock exchange in the HSE06 hybrid
functional. Detailed considerations concerning numerical and theoretical treatment of the mixing parameter
$a$ in terms of ${\epsilon^{-1}_{\infty}}$ can be found in Refs.~\cite{Franchini2012,Moussa2012}. Based on
that, we could also mention that due to the relation $r_{\textrm{TF}}>>(2/\mu)$, which is common to
ion-covalent insulators, such a parametrization is generally safe for tuning the optimal $a$ (here
$r_{\textrm{TF}}$ is the Thomas-Fermi screening length and $\mu$ characterizes the range separation in the
HSE06 hybrid functional). Evaluation of the mixing coefficient $a$ was carried out according to the
iterative procedure as follows: First, lattice relaxation has been performed and dielectric properties was
studied within the PBE-GGA. Then the macroscopic dielectric tenzor was evaluated using the density
functional perturbation theory \cite{Kresse2006}. Taking these elements as initial guess for $a$ via
${\epsilon^{-1}_{\infty}}$ (${\epsilon_{\infty}}=(2{\epsilon_{\infty}^{xx}}+{\epsilon_{\infty}^{zz}})/3$),
improvements to the macroscopic dielectric matrix was obtained by using $a$ as argument in the modified
HSE06 hybrid functional. The calculated ${\epsilon_{\infty}}$ and $a$ are provided in
Table~\ref{tab:Table_eps}. In these calculations the local field (LF) effects have been included into
consideration. The subsequent self-consistent electronic structure calculations were carried out using the
relaxed lattice geometries within the HSE06 functionals and the mixing parameters $a$ of
Table~\ref{tab:Table_eps}.

\begin{table}[htbp]
\caption{\label{tab:Table_eps} Calculated macroscopic dielectric matrix and fraction of the Fock exchange
mixing coefficient $a$.} \vspace{0.25cm}
\begin{tabular}{llcccc}
\hline
\multicolumn{2}{l}{\textrm{Compound}}{} & $\epsilon_{\infty}^{xx}$ & $\epsilon_{\infty}^{zz}$ & $\epsilon_{\infty}$ & $a$ \\
    \hline
\ce{Mg(OH)2} & PBE:   & 2.63 & 2.60 & 2.62 &        \\
             & HSE06: & 2.34 & 2.38 & 2.35 & 0.425  \\
\ce{Ca(OH)2} & PBE:   & 2.73 & 2.52 & 2.66 &        \\
             & HSE06: & 2.36 & 2.26 & 2.33 & 0.429  \\
\hline
  \end{tabular}
\end{table}

\begin{figure*}[!htb]
\includegraphics[width=1.0\textwidth,keepaspectratio=true]{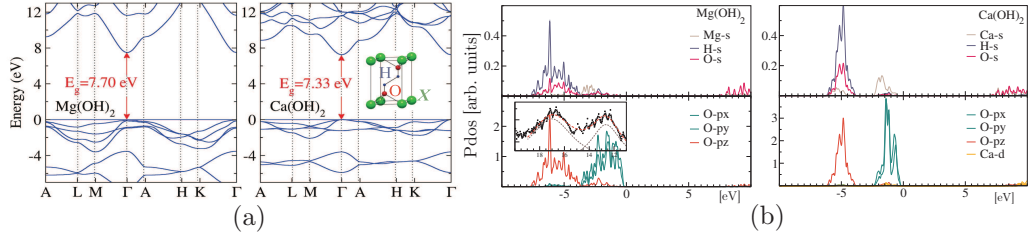}%
\caption{\label{fig:Fig_1} (Color online) Band structure and PDOS for \ce{\textit{X}(OH)2}. Topmost VB is
set to zero. PDOS for \ce{Mg(OH)2} has been compared to the experimentally measured XPS spectra
\cite{Haycock1978} plotted in the inset by ($\cdot\cdot\cdot$) and interpolated (solid lines).}
\end{figure*}

Fig.~\ref{fig:Fig_1} displays band structure as well as orbital and site projected density of states
(PDOS). The energies of three valence bands (VB) and two conduction bands (CB) as well as the symmetry
classification of the bands at the ${\Gamma}$ point are listed in Table~\ref{tab:Table_Gsymm}. Analysis
shows that \ce{\textit{X}(OH)2} are direct band gap materials with the VB maximum and CB minimum located
at the $\Gamma $-point. Topmost part of the VB shows splitting into two oxygen-derived subbands. One is
twofold degenerate $p_x, p_y$ bands and the other one is non-degenerate $p_z$ bands denoted by $E_{v1}$,
$E_{v2}$ and $E_{v3}$, respectively. Double degeneracy of the $p_x, p_y$ orbitals is lifted along the
$\Gamma$-$K$ and $\Gamma$-$M$ directions, but it conserves along the $\Gamma$-$A$ direction.

\begin{table}[htbp]
\caption{\label{tab:Table_Gsymm} Energies (in eV) of the VB and CB edges near to the fundamental band gap
and their symmetry classification at the ${\Gamma}$~point (in curly brackets). }
\vspace{0.25cm}
\begin{tabular}{lcc}
\hline
Band  & \ce{Mg(OH)2}: ${\Gamma}$ & \ce{Ca(OH)2}: ${\Gamma}$  \\
    \hline
$E_{v3}$  & $-3.41 (A_{2u}+A_{1g}) $ & $-3.61 (A_{2u}+A_{1g}) $ \\
$E_{v2}$  & $-0.79 (E_{u}) $         & $-1.21 (E_{u}) $         \\
$E_{v1}$  & $0 (E_{u})       $       & $0 (E_{u}) $             \\
$E_{c1}$  & $7.70 (A_{1g}+A_{2u}) $  & $7.33 (A_{1g}+A_{2u}) $  \\
$E_{c2}$  & $11.84 (A_{1g}+A_{2u})$  & $11.56 (A_{1g}) $        \\
\hline
  \end{tabular}
\end{table}

It follows from Fig.~\ref{fig:Fig_1} that contribution of $s$-, $p$-, and $d$-states into PDOS
significantly differ each from other. One can see that the \ce{Mg}/\ce{Ca} atoms donate their valence
$s$-electrons to the \ce{O} atoms which results in dominance of the oxygen 2$p$ orbitals and partially the
hydrogen $s$ orbitals in the VB, whereas the lowest empty states of the CB belong mainly to
\ce{Mg}/\ce{Ca} and \ce{O} $s$-type orbitals. They are also responsible for the fundamental band gaps of
$7.70$ and $7.33$ eV for \ce{Mg(OH2)} and \ce{Ca(OH2)}, respectively, corresponding to transitions from
\ce{O} $2p_x$ and $2p_y$ states.

Strong $sp_z$ hybridization between \ce{H} and \ce{O} orbitals repels drastically the $2p_z$ bands toward
lower energies. Electron localization function analysis has shown that such asymmetry in local angular
characters of $2p$-states is rooted in overlapping two chemical interactions as follows: the electron
transferred to the O atom along \ce{Mg-O}/(\ce{Ca-O}) connections is more likely to stay away from the
charge-transfer channel to support a classical ionic bond, i.e. an electrostatic interaction of the
anionic hydroxyl group with the metal cation in the $xy$ plane. The other electron, which was contributed
by the hydrogen atom, remains located inside the \ce{OH$^-$} ion to be employed mostly in covalent
\ce{O-H} bond along $z$ axis. This leads to an embedding of covalent bonding into the elemental framing of
the ionic structure, stipulating thus the splitting of an angular character of the oxygen electronic
states.

Calculated PDOS for \ce{Mg(OH)2} projected on \ce{O} atoms [Fig.~\ref{fig:Fig_1}] are in a good agreement
with the experimentally measured X-ray photoelectron spectra (XPS) \cite{Haycock1978} in the energy range
$0$-$8$~eV. This allows us to ascribe the two adjacent XPS peaks located at $12.7$~eV and $17.0$~eV to the
\ce{O} ($2p_x,2p_y$) and $2p_z$ states, respectively. It is interesting to note that the energy difference
between midpoints of the experimentally measured peak widths match well with the results of our
calculations.

From the electron band picture one can reveal two main channels of electric-dipole transitions differing
significantly in character: the one stems from the oxygen ($2p_x, 2p_y$) orbitals providing the direct
band gap transitions, whereas the second forms the transition channel that is associated with the
hybridized $2p_z$ and $s$ orbitals and becomes pronounced at higher exciting energies. Due to the lack of
any noticeable $d$-related contributions near the fundamental absorption edge one can suggest that the
polarization dependencies of the dipolar part of the target optical spectra in this region will mainly
differ according to dipolar selection rules connected with the ($2p_x, 2p_y$) and $p_z$ oxygen states,
respectively. Correspondingly, in the region of the higher energies where \ce{Ca} $3d$-related states
become sufficiently large, the dipole transition matrix element will begin additionally to select these
states correcting thus the polarization dependencies from $p$-type orbitals.

\begin{figure}[!htb]
\includegraphics[width=1.00\columnwidth,keepaspectratio=true]{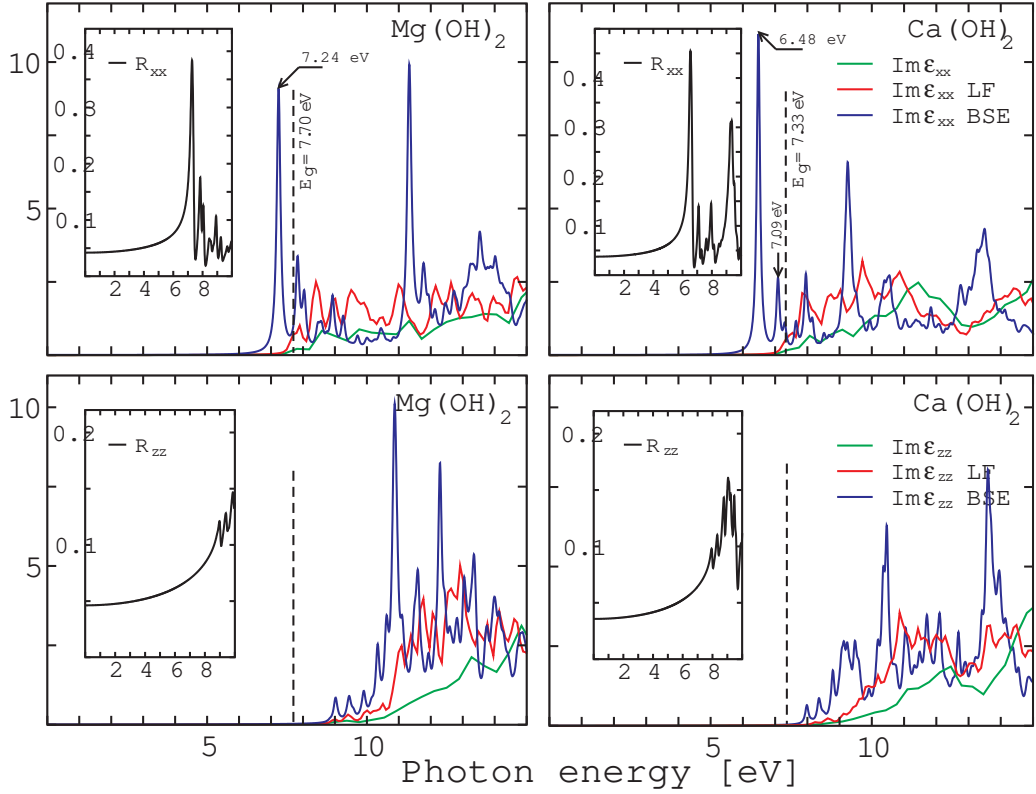}%
\caption{\label{fig:Fig_2}(Color online) The imaginary part of the macroscopic dielectric constants
calculated in the independent-particle approximation (i) without and (ii) with inclusion of LF effects,
and  (iii) within the ${\rm G_{3}W_{0}-BSE}$ approach. The latter method has been used for calculation of
the reflectivity spectra $R(\omega)$ plotted in the inset.}
\end{figure}

We have studied optical properties of \ce{\textit{X}(OH)2} for the photon energy range $0$-$15$~eV within
the frequency-dependent GW$_0$ method as implemented in VASP
\cite{Kresse2006GW,Kresse2007GW_1,Kresse2007GW_2,Kresse2007GW_3}. Our G$_3$W$_0$ calculations have been
performed by using HSE06 electronic-structure descriptions as a starting point. Figure~\ref{fig:Fig_2}
displays the imaginary part of the macroscopic dielectric constant. Analysis shows that
${\Im}{\epsilon_{xx}(\omega)}$ and ${\Im}{\epsilon_{zz}(\omega)}$ differ from each other significantly. It
indicates anisotropy in the optical properties of \ce{\textit{X}(OH)2}. Strong bound and resonance
exciton-related peaks are seen in the ${\Im}{\epsilon_{xx}(\omega)}$ spectra inside the band gap and in
the CB, respectively. One of the experimentally measurable parameters allowing to confirm existence
excitons is reflectivity spectra $R(\omega)$ plotted in the inset of Fig.~\ref{fig:Fig_2} where one can
see strong excitonic peaks.

In order to explain the anisotropy of the optical properties, it could be helpful to look at
Fig.~\ref{fig:Fig_1} in terms of the orbital splitting of the oxygen valence states, presuming that just
the longitudinal components of $\epsilon_2(\omega)$ refer to the fundamental absorption edge. Next we
recall a fact well-known for polar semiconductors and, especially, for insulators that the dielectric
function calculated in the independent-particle approximation is not so accurate as that obtained with
accounting for the LF effects. From Fig.~\ref{fig:Fig_2} we can see the distinct difference between these
two types of calculations -- the account for the LF effects changes both the peak positions and their
intensities in the absorption spectra of the hydroxides significantly. A further point to be made is that
the peak positions and mostly the intensities of the electron transitions observed in the ${\rm
G_{3}W_{0}-BSE}$ optical spectra of Fig.~\ref{fig:Fig_2} are, due to effects of the strong electron-hole
coupling, substantially changed in comparison with the noninteracting ones. This implies formation of a
bound exciton which corresponds to a new high spectral peak lying lower in energy than the fundamental
band gap. Theoretical predictions of these gaps along with the main characteristics of the exciton branch
of the optical spectra are summarized in Table~\ref{tab:Table_gaps}. Analysis shows that the ${\rm
G_{3}W_{0}}$ quasiparticle band gaps look comparable (slightly larger) with those obtained within the
HSE06 functional. This agreement gives us a reasonable confirmation that for \ce{\textit{X}(OH)2} the
HSE06 wave functions and eigenvalues represent a good starting point for ${\rm GW}$ numerical procedures.

\begin{table}[htbp]
\caption{\label{tab:Table_gaps} Calculated fundamental band gaps, $E_g$, and exciton peaks (in eV) in the
optical spectra. MBJ stands for the band gap prediction performed with the modified Becke-Johnson
potential \cite{Becke2006,Tran2009}.} \vspace{0.25cm}
\begin{tabular}{lcccccc}
\hline
Compound & PBE & HSE06m & MBJ & ${\rm G_{3}W_{0}}$ & $\rm E_{1}^{exc}$ & $\rm E_{2}^{exc}$ \\
    \hline
\ce{Mg(OH)2} & $3.83$ & $7.70$ & $7.16$  & $8.26$ & $7.24$ & $7.83$ \\
\ce{Ca(OH)2} & $3.76$ & $7.33$ & $7.20$  & $7.55$ & $6.48$ & $7.09$\\
\hline
  \end{tabular}
\end{table}

Our last step is proper interpretation of excitonic signatures seen in the optical spectra of
Fig.~\ref{fig:Fig_2}. Analysis of PDOS [Fig.~\ref{fig:Fig_1}] shows that mainly the combination of oxygen
and metal are responsible for the excitons possessing largest binding energy. In terms of characters of
the electronic states centered either on oxygen anion or metal cation the following two assignments of
one-photon vertical transitions to low-lying excited states can be made: (i) the intra-"molecular"
transition \ce{O^2-}${\rightarrow}${\ce{O^2-}} with the excited configuration
$2p_{z}^{2}2p_{x,y}^{3}3s^{1}$, since \ce{OH-} $sp_z$ bonding orbitals are situated significantly below,
low-lying excited states correspond to planar electronic excitations relating to oxygen of the hydroxyl
anion, and (ii) the transition \ce{O^2-}+ \ce{\textit{X}^{2+}}${\rightarrow}${\ce{O-}}+{\ce{\textit{X}+}}
with the excited configurations $2p^5$ and $3p^{6}4s^{1}$ for \ce{O} and \textit{X}, respectively; this
nonlocal (off-site) transition deals with the induced backward charge transfer along \ce{Mg-O} connection
and is associated with the partial decrease of the $p_{x,y}$ electron density on the oxygen and the
proportionate increase of the $s$ electron density on the metal cation. Thus, the central spectral feature
is splitting of the excitonic branch of Fig.~\ref{fig:Fig_2} between two prominent peaks of different
intensity, which can be regarded as follows: the larger peak at $7.24$ eV ($6.48$~eV for \ce{Ca(OH)2})
corresponds to the strong molecular component of the electron-hole state localized at the oxygen of the
hydroxyl anion, whereas the origin of the smaller one at $7.83$~eV ($7.09$~eV) is suggested to be much
closer to the electron-hole pair separated on the nearest-neighbors \ce{Mg-O}.

In order to explain the strong excitonic binding in the hydroxides we refer mainly to a rigorous physical
picture established for ion-covalent compounds -- the more the electronic charge is localized, the greater
strength many-body contributions exert \cite{Hanke1984,Su-Huai_exciton_2013}. In this context, aside from
the standard feature of the weak screening in the hydroxides we would emphasize the key role of the
unoccupied \ce{O^{2-}} \textit{s}-type orbitals located in the bottommost CB and hybridized with metal
cation $s$ orbitals, which, according to Ref.~\cite{Su-Huai_exciton_2013}, significantly strengthen the
Coulomb and exchange parts of the electron-hole interaction.

\section{Summary}
\label{summary}

In summary, we have presented the first-principles description of the electronic structure and optical
properties of multifunctional hydroxides \ce{\textit{X}(OH)2} by using DFT calculations with HSE06 hybrid
functional and many-body calculations with GW-BSE. Within the frameworks of the combined theoretical
approach we have found an intimate connection between the crystal-chemical properties and the relevant
features of the electronic spectra. With respect to the optoelectronic properties of \ce{Mg(OH)2} and
\ce{Ca(OH)2} this has allowed us to provide rationalization for a strong Coulomb attraction between an
excited electron and a hole and thereby to predict for the first time the existence of excitonic states
possessing large binding energy. Imaginary part of the dielectric function and reflectivity spectra show
strong peaks corresponding to the bound excitons. The hexagonal lattice environment has an ultimate effect
on the excitonic states making their structure highly anisotropic and directly dependent on the
polarization of the incident light in such a way that it can be realized only in the $a-b$ plane. We have
identified and analyzed the origin of the excitons, which can be attributed to strong localization of the
hole and electron to oxygen $2p_x, 2p_y$ occupied states as well as to oxygen and earth metal $s$ empty
states, respectively. Our results are expected to initiate further experimental and theoretical studies of
\ce{Mg(OH)2} and \ce{Ca(OH)2}. Also our theoretical results can serve as solid input data for developing
different experimental designs and synthesis based on the \ce{\textit{X}(OH)2} hydroxides.

\section*{Acknowledgement}
This work has received financial and supercomputing support from the Research Council of Norway through
the ISP project 181884. This work was also supported by the European Union through the European Regional
Development Fund (Centre of Excellence "Mesosystems: Theory and Applications", TK114) and by the Estonian
Science Foundation grant No 7296. The authors wish to thank Professor Y. Galperin, University of Oslo,
Oslo, Norway and Professor M. Ganchenkova, National Research Nuclear University, Moscow, Russia for
critical reading of the manuscript and useful comments as well as Dr. \O. Nordseth, Institute for Energy
Technology, Kjeller, Norway for practical help.


\newpage
\bibliographystyle{elsarticle-num}


\providecommand{\noopsort}[1]{}\providecommand{\singleletter}[1]{#1}%

\end{document}